\documentclass[aps,prl,amsmath,amssymb,twocolumn]{revtex4}
\usepackage{epsfig,amssymb}
\usepackage{graphicx}
\usepackage{dcolumn}
\usepackage{bm}
\usepackage{dcolumn}
\usepackage{multirow}
\begin{document}
\title{Physical origin of satellites in photoemission of doped graphene: an \emph{ab initio} GW
  plus cumulant study}

\author{Johannes~Lischner, Derek~Vigil-Fowler, and Steven~G.~Louie}

\affiliation{Department of Physics, University of California,
  Berkeley, California 94720, USA, and Materials Sciences Division,
  Lawrence Berkeley National Laboratory, Berkeley 94720, USA.}

\begin{abstract}
  We calculate the photoemission spectra of suspended and epitaxial
  doped graphene using an \emph{ab initio} cumulant expansion of the
  Green's function based on the GW self-energy. Our results are
  compared to experiment and to standard GW calculations. For doped
  graphene on a silicon carbide substrate, we find, in contrast to
  earlier calculations, that the spectral function from GW only does
  not reproduce experimental satellite properties. However, \emph{ab
    initio} GW plus cumulant theory combined with an accurate
  description of the substrate screening results in good agreement
  with experiment, but gives no plasmaron (i.e., no extra well-defined
  excitation satisfying Dyson's equation).
\end{abstract}

\maketitle

\emph{Introduction}.---The isolation of graphene \cite{Geim}, an
atomically thin layer of carbon, in 2004 has sparked much interest and
an enormous amount of scientific work. Among the most fascinating
properties of this extraordinary material is its electronic structure:
the low energy valence and conduction bands form two Dirac cones in
the Brillouin zone causing electrons in graphene to behave like
massless relativistic particles with chiral character.

Recently, several experimental groups probed the low-energy electronic
structure of doped graphene on a silicon carbide (SiC) substrate using
angle-resolved photoemission spectroscopy (ARPES)
\cite{Rotenberg1,Rotenberg2,Lanzara}. Surprisingly, the experimental
spectra did not exhibit simple Dirac cones, but indicated a more
complicated electronic structure. There has been much discussion
regarding the interpretation of the observed findings: while Lanzara
and coworkers \cite{Lanzara} concluded that the SiC substrate induces a
band gap in graphene, Rotenberg and coworkers
\cite{Rotenberg1,Rotenberg2} argued that the observed electronic
structure is intrinsic to graphene and is due to a \emph{plasmaron}
excitation caused by strong electron-carrier plasmon coupling.

Both interpretations were backed by theoretical findings: Kim et
al. \cite{Kim} carried out density-functional calculations of graphene
on a reconstructed SiC surface and found that the dangling bonds of
the substrate bind to the graphene layer breaking the symmetry of the
graphene sheet and causing a band gap to open. On the other hand,
Polini et al. \cite{Polini} and Hwang and Das Sarma \cite{DasSarma}
carried out Green's function calculations on the Dirac Hamiltonian
(model system with two linear bands) using the GW approximation
\cite{HedinBook} to the self-energy and a simplified description of
the substrate. These calculations found strong low-energy plasmaronic
bands which obey the experimentally observed scaling law as a function
of doping, and Bostwick \emph{et al.} reported agreement of theory with
experiment after introducing an adjustable parameter attributed to the
SiC substrate screening\cite{Rotenberg2}.

The agreement of GW calculations and experiment for satellite
properties in doped graphene is surprising since it is known that for
bulk solids, such as silicon or potassium
\cite{Langreth,HedinAryasetiawan,Guzzo}, electron correlations
\emph{beyond the GW approximation} are needed to accurately describe
plasmon satellites. For bulk solids, the GW plus cumulant (GW+C)
approximation \cite{HedinAryasetiawan,Guzzo}, which includes
significant vertex corrections, provides an accurate description of
plasmon satellites, but does not find plasmarons. However, no
application of the GW+C theory to a two dimensional system, such as
doped graphene with its characteristic low-energy carrier plasmon, has
been reported so far. Also, in contrast to bulk solids, the carrier
plasmon dispersion in graphene is \emph{tunable} and is both substrate
and doping dependent.

In this Letter, we investigate satellites in photoemission spectra of
doped suspended graphene and doped graphene on a SiC substrate using
the \emph{ab initio} cumulant expansion of the Green's function based
on the GW approximation to the self-energy. We compare our findings
with those from \emph{ab initio} GW calculations and also with GW
calculations based on the model Dirac Hamiltonian. As a benchmark, we
also calculated \emph{ab initio} GW+C photoemission spectra for
silicon.

\emph{Methods}.---The photoelectron current in an ARPES experiment
with monochromatic photons of frequency $\nu$ and polarization
$\hat{e}_\nu$ is given by \cite{damascelli}
\begin{equation}
  I(\bm{k},\omega,\hat{e}_\nu,\nu)= 
  \sum^{\text{occ}}_n I_0(n,\bm{k},\omega,\hat{e}_\nu,\nu)
  f(\omega) A_{n\bm{k}}(\omega),
\end{equation}
where $\bm{k}$ and $\omega$ are the momentum and binding energy of an
electron in band $n$, $f(\omega)$ is the Fermi-Dirac distribution and
$I_0$ includes the absorption cross section of the incident
photons. Also, $A_{n\bm{k}}(\omega)=1/\pi |\text{Im}
G_{n\bm{k}}(\omega)|$ denotes the spectral function with
$G_{n\bm{k}}(\omega)$ being the interacting one-particle Green's
function.

Usually, $G_{n\bm{k}}(\omega)$ is obtained by solving Dyson's equation
$G^{-1}_{n\bm{k}}(\omega) =
G^{-1}_{0,n\bm{k}}(\omega)-\Sigma_{n\bm{k}}(\omega)+V^{xc}_{n\bm{k}}$
with $G_{0,n\bm{k}}(\omega)$ and $V^{xc}_{n\bm{k}}$ denoting a
mean-field Green's function and exchange-correlation potential,
respectively, and $\Sigma_{n\bm{k}}(\omega)$ is the self-energy. In
this work, we employ the \emph{ab initio} GW approach to the
self-energy \cite{HedinBook,LouieHybertsen}.

While describing quasiparticle properties in many materials with high
accuracy \cite{LouieHybertsen}, the GW approximation is less reliable
for satellite properties \cite{Langreth,HedinAryasetiawan,Guzzo}: for
the spectral function of a core electron interacting with plasmons, GW
predicts a single satellite instead of a satellite series with
decreasing spectral weight and also greatly overestimates the binding
energy of the satellite structures. The cumulant expansion
\cite{HedinAlmbladh,Hedin,HedinAryasetiawan} of $G_{n\bm{k}}(\omega)$
cures these deficiencies by including significant vertex corrections
beyond GW: it provides the \emph{exact} solution for a core electron
interacting with plasmons \cite{Langreth}. In the cumulant approach,
the Green's function for a hole is expressed as
\begin{align}
  G_{n\bm{k}}(t) = i \Theta(-t) e^{-i\epsilon_{n\bm{k}}t + C_{n\bm{k}}(t)},
  \label{Gcumulant}
\end{align}
where $\epsilon_{n\bm{k}}$ denotes the mean-field orbital energy and
$C_{n\bm{k}}(t)$ denotes the cumulant. This expression for the Green's
function is obtained after the first iteration of the self-consistent
solution of its equation of motion assuming a simple quasiparticle
form for the starting guess \cite{HedinAlmbladh}.  

The cumulant can be separated into a quasiparticle part
$C^{qp}_{n\bm{k}}(t)$ and a satellite part $C^{sat}_{n\bm{k}}(t)$
given formally in terms of the self-energy by (for $t<0$)
\begin{align}
  C^{qp}_{n\bm{k}}(t) &= -it \Sigma_{n\bm{k}}(E_{n\bm{k}}) + 
  \frac{\partial \Sigma^h_{n\bm{k}}(E_{n\bm{k}})}{\partial \omega} \label{Cqp}\\
  C^{sat}_{n\bm{k}}(t) &= \frac{1}{\pi} \int_{-\infty}^{\mu} d\omega 
  \frac{\text{Im}\Sigma_{n\bm{k}}(\omega)}{(E_{n\bm{k}}-\omega-i\eta)^2}
  e^{i(E_{n\bm{k}}-\omega)t},
  \label{Csat}
\end{align}
where $\mu$ denotes the chemical potential, $\eta$ is a positive
infinitesimal,
$E_{n\bm{k}}=\epsilon_{n\bm{k}}+\Sigma_{n\bm{k}}(E_{n\bm{k}})-V^{xc}_{n\bm{k}}$
is the quasiparticle energy and $\Sigma^h_{n\bm{k}}(\omega)$ is
defined through the relation
\begin{align}
\Sigma^h_{n\bm{k}}(\omega)=\frac{1}{\pi}
\int_{-\infty}^\mu d\omega'
\frac{ \text{Im}\Sigma_{n\bm{k}}(\omega')}{\omega'-\omega-i\eta}.
\label{Sigmah}
\end{align}

For a given level of approximation for $\Sigma$, the cumulant theory
yields an improved Green's function through
Eqs.~(\ref{Gcumulant}-\ref{Sigmah}). In the present study, $\Sigma$ is
obtained from \emph{ab initio} GW theory \cite{LouieHybertsen} which
is known to describe quasiparticle properties in Si, C and related
materials accurately thus providing a good starting point for the
cumulant theory.

\emph{Silicon}.---To benchmark the accuracy of the \emph{ab initio}
GW+C approach, we first applied it to silicon where accurate
angle-integrated photoemission data is available \cite{Guzzo}.  We
carried out one-shot full frequency G$_0$W$_0$ calculations for
$\Sigma$ using the BerkeleyGW package\cite{BGWpaper} [see
Supplementary Materials (SM) for details] and then evaluated the
cumulant spectral function.

\begin{figure}
  \includegraphics[width=8.cm]{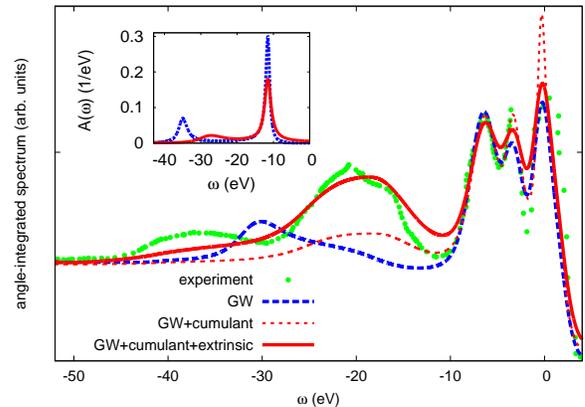} 
  \caption{Comparison of theoretical results and experiment
    data\cite{Guzzo} for the angle-integrated photoemission spectrum
    of silicon. The inset shows the spectral function of the lowest
    valence band at the $\Gamma$-point ($\Gamma_{1v}$) in silicon from
    \emph{ab initio} GW (dashed blue curve) and \emph{ab initio} GW+C
    (solid red curve) theories. The energy is measured from the top of
    the valence band}
  \label{fig:silicon}
\end{figure}

The inset of Fig.~\ref{fig:silicon} compares the spectral functions
from \emph{ab initio} GW and \emph{ab initio} GW+C theory for the
lowest valence band at the $\Gamma$ point of silicon. While the GW
spectral function exhibits a strong plasmaron peak due to an additonal
spurious solution of Dyson's equation
\cite{Langreth,HedinAryasetiawan,Guzzo} (see SM) separated by $\sim
23$~eV from the quasiparticle peak, the GW+C spectral function shows a
more shallow peak separated by $\sim 16$~eV from the quasiparticle
peak. This separation agrees well with the experimental plasmon energy
of $16.6$~eV \cite{Ehrenreich} in silicon indicating a much weaker
interaction between electrons and plasmons than predicted by GW.

Figure~\ref{fig:silicon} shows the angle-integrated photoemission
spectrum from the \emph{ab initio} GW+C theory and compares it to the
\emph{ab initio} GW result and experiment. We have included matrix
element effects, the inelastic background and extrinsic plasmon losses
similar to the work of Guzzo and coworkers \cite{Guzzo} (see SM). Our
full frequency \emph{ab initio} GW+C theory reproduces accurately the
location, height and width of the first plasmon peak and it does not
show a significant second peak.

\emph{Doped Graphene}.---In contrast to silicon where the plasmons are
high-energy excitations, doped graphene exhibits a low-energy,
dispersive plasmon branch due to added carriers from doping with a
characteristic 2D square root of q dispersion relation
\cite{Rotenberg2,Polini,DasSarma}, the so-called carrier plasmon.  Due
to the special dispersion relations of the carrier plasmons and of the
Dirac fermions, the self-energy of doped graphene exhibits very sharp
features in ($\bm{k}$,~$\omega$)-space \cite{Polini,DasSarma}
necessitating \emph{extremely} fine k-point and frequency grids for
convergent results: our calculation of $\Sigma_{n\bm{k}}(\omega)$ and
$A_{n\bm{k}}(\omega)$ employs a full frequency sampling at 0.05~eV
intervals and a $1440\times 1440\times 1$ k-point grid (see SM).

\begin{figure}
  \includegraphics[width=8.cm]{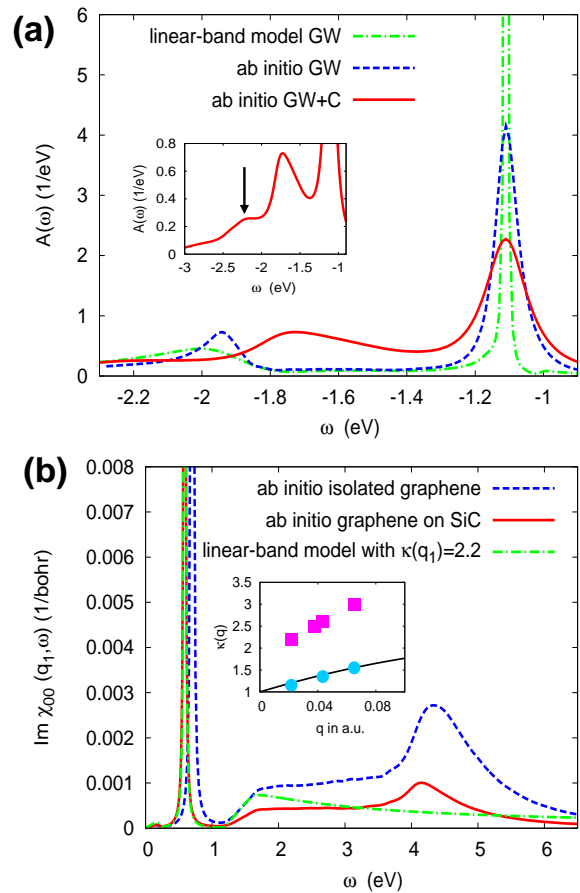} 
  \caption{(a): Comparison of \emph{ab initio} GW+C, \emph{ab initio}
    GW, and linear-band model GW spectral functions at the Dirac point
    for isolated graphene with electron doping corresponding to a
    charge density $n=-12.4\times 10^{13}$~cm$^{-2}$. Energies are
    measured relative to the Fermi energy. The inset shows an
    additional shoulder-like feature (indicated by arrow) in the
    \emph{ab initio} GW+C spectral function. (b): Imaginary part of
    the head of the interacting susceptibility (in a supercell
    calculation $\text{Im}\chi_{00}(\bm{q},\omega)=\text{Im}
    \epsilon^{-1}_{00}(\bm{q},\omega)/v(\bm{q})/n_g$ with $v(\bm{q})$
    denoting the truncated Coulomb interaction and $n_g$ the density
    of graphene sheets per unit length) for isolated graphene and for
    graphene on a SiC substrate. Note that
    $\text{Im}\chi_{00}(\bm{q},\omega)$ is proportional to the
    dynamical structure factor $S(\bm{q},\omega)$ \cite{Sensarma}. We
    also show the result from the linear-band model with a background
    dielectric of $\kappa(q_1)=2.2$. The results are for $n=-5.9\times
    10^{13}$~cm$^{-2}$ and $q_1=0.022$~a.u.. The inset shows the
    background dielectric function $\kappa(q)$ used in the strictly 2D
    linear-band model for isolated graphene (solid circles), graphene
    on SiC (solid squares) and the cRPA result for isolated graphene
    obtained by Wehling \emph{et al.} (solid line)\cite{bluegel}.}
  \label{fig:graphene_dirac}
\end{figure}

Figure~\ref{fig:graphene_dirac}(a) shows the \emph{ab initio} GW+C
spectral function at the Dirac point of isolated graphene with
electron doping resulting in a charge density $n=-12.4\times
10^{13}$~cm$^{-2}$. Also shown are the \emph{ab initio} GW spectral
function and the GW spectral function from the linear-band
model\cite{Polini,DasSarma}. (In the linear-bands model, graphene is
described by the Dirac Hamiltonian with a bandstructure consisting of
only two linear bands at the K and K' points of the Brillouin zone.)
For the linear-band model, we included screening contributions from
all other bands by adding the analytically parametrized cRPA
dielectric function calculated within DFT by excluding the p$_z$
bands, obtained by Wehling and coworkers (Eq.~(3) of
Ref.\cite{bluegel}), to the calculated dielectric function of the
linear-band model \cite{DasSarmaDiel,Stauber}. We verified the
accuracy of the cRPA dielectric function of Wehling \emph{et al.} by
comparing it to the additional background dielectric constant
$\kappa(q)$ necessary to match the carrier plasmon peak of the
imaginary part of the interacting susceptibility from the linear-band
model to the corresponding peak in the \emph{ab initio} calculation
[see inset of Fig.\ref{fig:graphene_dirac}(b)].

Despite the simplification of the graphene band structure in the
linear-band model, the resulting GW spectral function is similar to
the \emph{ab initio} GW result \emph{provided} that the above
$\kappa(q)$ is used: both theories give a strong quasiparticle peak
and a smaller satellite peak. The separation of the peaks is similar,
0.83~eV in \emph{ab initio} GW and 0.89~eV in the linear-band
model. However, the shapes of the features differ noticeably: in the
\emph{ab initio} GW theory, the quasiparticle peak is much broader
than in the linear-band model, while the satellite peak is sharper
revealing the importance of a realistic bandstructure for high doping
levels. The GW approximation leads to a spurious plasmaron solution of
Dyson's equation giving rise to the peak near -2~eV in Fig.~2(a) (see
SM). The inclusion of higher order electron interaction effects in the
\emph{ab initio} GW+C calculations show important changes in the
spectral function in Fig.~\ref{fig:graphene_dirac}: i) there is a
significant shift of the amplitude from the quasiparticle peak to the
satellite peak, and ii) the peak separation is reduced to 0.62~eV, a
change of $\sim 30$~percent. The inset in
Figure~\ref{fig:graphene_dirac}(a) shows that the \emph{ab initio}
GW+C spectral function also exhibits an additional shoulder-like
structure at $\omega\sim -2.3$~eV resulting from the coupling of the
hole to \emph{two} plasmons. So far no experimental ARPES spectrum
with sufficient accuracy to observe satellite features has yet been
reported for suspended graphene.

\begin{figure}
  \includegraphics[width=8.cm]{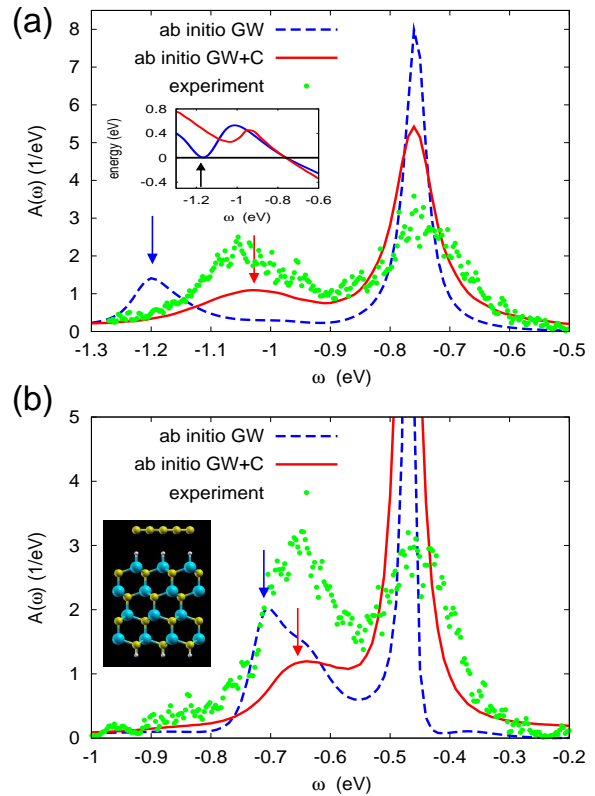} 
  \caption{Comparison of \emph{ab initio} GW+C and GW spectral
    function (in 1/eV) for graphene on SiC at the Dirac point with
    experiment (experimental spectra are given in counts/second in
    arbitrary units) for (a) $n=-5.9\times 10^{13}$~cm$^{-2}$ and (b)
    $n=-1.5\times 10^{13}$~cm$^{-2}$. The arrows indicate the
    satellite position in the theoretical curves. The inset of (a)
    shows $\text{Re}\delta\Sigma(\omega)-\omega+\epsilon_{LDA}$ with
    $\delta\Sigma=G^{-1}_0-G^{-1}$ for GW (blue curve) and GW+C (red
    curve) as a function of $\omega$. The arrow denotes the plasmaron
    solution in GW, which does not exist in GW+C. The inset of (b)
    shows the geometry of graphene on a hydrogen-terminated 4H-SiC
    (0001) surface: carbon atoms are brown, silicon atoms are blue and
    hydrogens are white. }
  \label{fig:graph_SiC}
\end{figure}

Bostwick and coworkers had carried out ARPES experiments on
hydrogen-intercalated epitaxial graphene grown on SiC(0001)
\cite{Rotenberg2,Riedl}. To include the screening effect of the
substrate we model the substrate by a 10-atomic-layer thick slab of
hydrogen-terminated 4H-SiC(0001) employing the $\sqrt{3}\times
\sqrt{3}R30^{\circ}$ SiC surface unit cell
\cite{Varchon,Pankratov,Krukowski} [see inset of
Figure~\ref{fig:graph_SiC}(b)] resulting in a supercell geometry with
neighboring graphene sheets separated by 54~a.u.. To determine the
SiC-graphene distance we carried out density-functional theory
calculations with the empirical van der Waals correction of Grimme
\cite{Grimme} and obtain a separation of $3.86~\AA$ which agrees well
with the findings of Soltys and coworkers \cite{Krukowski}. Because of
the passivation of the surface dangling bonds by hydrogen atoms, we
also do not find a surface induced band gap opening in the graphene
sheet \cite{Krukowski}.

We then carried out GW calculations on the graphene+substrate system
using a truncated Coulomb interaction \cite{BGWpaper}. Due to the
immense size of the system under consideration, we separately
calculate the contributions to the polarizability matrix from the
graphene sheet and the substrate before adding them up (see SM).

Figure~\ref{fig:graphene_dirac}(b) shows the imaginary part of the
interacting susceptibility of graphene on H-passivated SiC for
$n=-5.9\times 10^{13}$~cm$^{-2}$ at a small wave vector
$q_1=0.022$~a.u. and compares it to the isolated graphene result. The
substrate screening leads to a reduction of the carrier plasmon
frequency (the position of the sharp peak at $\sim
0.7$~eV). Figure~\ref{fig:graphene_dirac}(b) also shows that the
plasmon peak of the linear-band model coincides with the \emph{ab
  initio} result if the 2D bare Coulomb interaction is divided by a
background dielectric constant $\kappa(q_1)=2.2$ which now contains
both the intrinsic graphene screening processes discussed earlier and
screening contributions from the substrate. This deduced \emph{ab
  initio} value of $\kappa$ is much smaller than the value used by
Bostwick and coworkers \cite{Rotenberg2} who treated $\kappa$ as a
\emph{q-independent fitting parameter} and found that $\kappa=5.15$
(using the LDA Fermi velocity to convert the coupling strength
$\alpha_G=0.5$ to a background dielectric constant) gives the best
description of the experimental spectra within the GW approximation
only. We find that $\kappa$ increases with $q$, but remains much
smaller than the value needed by Bostwick and coworkers to fit
experiment [see inset of Fig.~\ref{fig:graphene_dirac}(b)].

Figure~\ref{fig:graph_SiC}(a) compares \emph{ab initio} GW+C and GW
spectral functions for graphene on SiC at the Dirac point with
experiment for $n=-5.9\times 10^{13}$~cm$^{-2}$. All spectral
functions exhibit a quasiparticle peak and a satellite: their
separation is 0.44~eV in GW and 0.27~eV in the GW+C theory. The
\emph{ab initio} GW+C theory agrees very well with the experimental
separation of 0.30~eV \cite{Rotenberg2}.  Also, the shape of the GW+C
spectral function resembles the experimental spectrum more closely
than the GW result. The inset of Fig.~\ref{fig:graph_SiC}(a) shows
that GW predicts an additional plasmaron solution to Dyson's equation.
In contrast, no plasmaron solution is found if the vertex-corrected
self-energy $\delta\Sigma_{GW+C}=G^{-1}_0-G^{-1}_{GW+C}$ is used.  We
attribute the remaining difference between the experimental spectrum
and the computed spectral function to extrinsic losses (i.e., emission
of plasmons by the photoexcited electron), which are not included in
our theory (see SM for a preliminary treatment of extrinsic losses).

Bostwick and coworkers have shown that the experimental spectral
functions exhibit an important scaling behavior as function of doping
\cite{Rotenberg2}: when the experimental quasiparticle-satellite
separation is divided by the Fermi level $E_F$ the resulting number is
independent of doping for a wide range of doping levels, a consequence
of the linear dispersion of electrons in graphene. To check whether
the \emph{ab initio} GW+C method reproduces this scaling behavior we
carried out calculations for $n=-1.5\times 10^{13}$~cm$^{-2}$. The
results are shown in Fig.~\ref{fig:graph_SiC}(b). Again, we find that
the separation of quasiparticle and satellite peaks is described
accurately by the \emph{ab initio} GW+C method.

In conclusion, we have carried out a first-principles study of doped
graphene on hydrogen-intercalated SiC, going beyong standard GW
calculations, which explains the experimental findings of Bostwick and
coworkers \cite{Rotenberg2}. The work reproduces the experimental
satellite properties without finding a plasmaron showing the
importance of an advanced treatment of electron correlations via the
\emph{ab initio} GW+C method and an accurate modelling of substrate
screening. We do not find a substrate induced band gap or a plasmaron
solution. We also explain the previously reported good fit of GW
calculations to experiment using an simplified modelling of the
substrate \cite{Rotenberg2}. In these calculations, the lack of
sufficient electron correlations at the GW level leads to an extra
solution to Dyson's equation which gives rise to an overestimation of
the quasiparticle-satellite separation. However, these calculations
also overestimate the effect of substrate screening resulting in
spurious agreement with experiment.

We acknowledge useful discussions with Allan MacDonald, Marco Polini,
Alessandro Principi, Matteo Guzzo, Felipe da Jornada, Manish Jain, and
Aaron Bostwick. We acknowledge support in our initial work formulating
the conceptual foundations of this Letter and partial postdoctoral
support for one of us (J.L.) from National Science Foundation Grant
No. DMR10-1006184.  We also acknowledge support in the computational
part of this project and for student support (D.V.F.)  from the
Director, Office of Science, Office of Basic Energy Sciences,
Materials Sciences and Engineering Division, U.S. Department of Energy
under Contract No. DE- AC02-05CH11231. D.V.F was also supported by a
Department of Defense (DoD) National Defense Science \&
Engineering Graduate Fellowship. Computational
resources have been provided by the DOE at Lawrence Berkeley National
Laboratory’s NERSC facility and by the NSF through XSEDE resources at
NICS.

\bibliography{paper}
\end{document}